\documentclass[12pt]{article}%
\usepackage{amsfonts}
\usepackage{amsmath,amssymb,amsthm,enumerate,epsfig,graphicx}
\usepackage{ifthen,latexsym,syntonly}
\usepackage{amsmath}
\usepackage{amssymb}
\usepackage{graphicx}
\usepackage{verbatim}
\usepackage{morefloats}
\usepackage{geometry}
\usepackage{soul,color}%
\usepackage{cprotect}
\usepackage{paralist}
\usepackage{float}
\usepackage{pdflscape}
\usepackage{booktabs}

\newtheoremstyle{exampstyle}
{3pt} 
{3pt} 
{\itshape} 
{} 
{\bfseries} 
{} 
{.5em} 
{} 

\newtheorem{theorem}{Theorem}

\newtheorem{remark}{Remark}

\theoremstyle{exampstyle} 
\theoremstyle{exampstyle} 
\theoremstyle{exampstyle} 
\theoremstyle{exampstyle} 

\usepackage{fancybox}
\usepackage{multirow}
\usepackage[figuresleft]{rotating}

\usepackage{endnotes}
\let\footnote=\endnote

\makeatletter
\renewcommand{\tablename}{{\bf Table}}
\renewcommand{\fnum@table}[1]{\normalfont\textbf{\tablename~\thetable}.~}

\renewcommand{\theenumi}{\Alph{enumi}}

 \makeatletter
 \renewcommand{\p@enumii}{\theenumi.}
 \makeatother

\makeatother
\usepackage{indentfirst} 

\makeatletter
\renewcommand{\figurename}{{\bf Fig.}}
\renewcommand{\fnum@figure}[1]{\normalfont\textbf{\figurename~\thefigure.}}
\makeatother

\makeatletter
\newif\ifrhsapp
\rhsappfalse

\newcommand{\@seccntformat@section}[1]{%
  \ifrhsapp
  Appendix
  \else
  \fi
  \csname the#1\endcsname.\quad
}

\newcommand*{\@seccntformat@subsection}[1]{%
  \csname the#1\endcsname.\quad
}

\newcommand*{\@seccntformat@subsubsection}[1]{%
  \csname the#1\endcsname.\quad
}

\let\@@seccntformat\@seccntformat
\renewcommand*{\@seccntformat}[1]{%
  \expandafter\ifx\csname @seccntformat@#1\endcsname\relax
    \expandafter\@@seccntformat
  \else
    \expandafter
      \csname @seccntformat@#1\expandafter\endcsname
  \fi
    {#1}%
}

\renewcommand{\subsection}{\@startsection
  {subsection}{2}{0mm}{-3.25ex \@plus -1ex \@minus -.2ex}{1.5ex \@plus.2ex}{\normalfont\large\itshape}}

\renewcommand{\subsubsection}{\@startsection
  {subsubsection}{2}{0mm}{-3.25ex \@plus -1ex \@minus -.2ex}{1.5ex \@plus.2ex}{\normalfont\itshape}}

\def\appendix{\par
  \setcounter{section}{0}%
  \setcounter{subsection}{0}%
  \rhsapptrue
  \renewcommand\thesection{\Alph{section}}%
}
\makeatother

\newtheoremstyle{query}%
{}{}
{\color{red}}
{}
{\sffamily\bfseries}{:}{12pt}
{}
\theoremstyle{query}

\begin{document}

\title{Sequential Sampling for  CGMY Processes via Decomposition of their Time Changes}
\author{\sc Chengwei Zhang,  Zhiyuan Zhang\thanks{Please send correspondence to: Zhiyuan Zhang, School of Statistics and Management, Shanghai University of Finance and Economics, No. 777 Guoding Road, Shanghai 200433, China. Tel: (+86) 021 6590 4159. E-mail address: zhang.zhiyuan@mail.shufe.edu.cn.}\\
{\footnotesize School of Statistics and Management, Shanghai University of Finance and Economics, Shanghai, China.}\\
}
\date{}
\maketitle
\begin{abstract}
We present a new and easy-to-implement sequential sampling method for CGMY processes  with either finite or infinite variation, exploiting the time change representation of the CGMY model and a decomposition of its time change. We find that the time change can be decomposed into two independent components. While the first component is a \emph{finite} \emph{generalized gamma convolution} process whose increments can be sampled by either the exact double CFTP (``coupling from the past'') method or an approximation scheme with high speed and accuracy, the second component can  easily be made arbitrarily small in the $L^1$ sense. Simulation results show that the proposed method is advantageous over two existing methods under a model calibrated to historical option price data.


\end{abstract}
{\it Keywords}: \hspace*{0mm}
sequential sampling; CGMY processes;  double CFTP; option pricing \\[0mm]

\newpage
\section{INTRODUCTION}
Jump processes have become increasingly popular in financial modeling since the seminal work of \cite{Merton1976} (see, e.g., \cite{Kou2002} and \cite{KouWang2004}). \cite{ContTankov} provides a comprehensive exposition of the use of jump processes in financial modeling. The CGMY model of \cite{CGMY2002} is one of the most popular jump processes.

CGMY processes are flexible pricing models that exhibit infinite activity and can be of either finite variation (i.e., with stability index $0<Y<1$) or infinite variation (i.e., with stability index $1\leq Y<2$). Since its inception, the CGMY model has found  success  in modeling both asset returns and option prices. \cite{CGMY2002} calibrated the CGMY model to both the (underlying) equity prices and option prices. Their empirical results show that price processes of most of the studied equities are of pure jump and infinite activity, and that both finite- and infinite-variation instances exist though the latter happens less frequently. The modeling flexibility of the CGMY model can be understood as follows. \cite{MadanYor} showed that a CGMY process can be represented as a Brownian motion time-changed by an independent subordinator that is usually referred to as the time change of the CGMY process. In fact, in the early 1970s, in the context of the modeling of asset prices, \cite{Clark1973} had already introduced the idea of time change, which can effectively capture such stylized empirical facts as fat-tailedness and skewness for the distribution of observed asset returns. \cite{AneGeman} later extended this idea to the modeling of the flow of market information by time changes in explaining the normality of observed asset returns. In option valuation, \cite{CGMY2003} found clear advantages of CGMY models endowed with stochastic volatility over other L\'evy models endowed with stochastic volatility in terms of reproducing the volatility skew pattern.


Nonetheless, one challenging problem with these otherwise appealing jump models is to find a sequential sampling (or path simulation) method that is pertinent to pricing path-dependent options.

In the finite-variation case (where the stability index $0<Y<1$), exact simulation methods are available. In this case, the density of the CGMY increment is an exponentially tilted density of a unilateral stable random variable, and therefore the standard rejection sampling method can be applied. However,  simple rejection sampling suffers from low acceptance rates in certain regions of  parameter space. To overcome this shortcoming of the simple rejection method, \cite{Devroye} developed an exact double rejection method with uniformly bounded complexity over all parameter ranges.


In the infinite-variation case (where the stability index $1\leq Y<2$), all available sampling methods entail approximations.  Utilizing the time change representation, \cite{MadanYor} developed a sequential simulation method for the CGMY model through sequentially sampling the increments of its time change. This method consists of two steps. In the first, one truncates and approximates the contribution of small jumps in the series representation for a $Y/2$-stable subordinator, following the approach of \cite{AsmussenRosinski}. Second, based on the approximate $Y/2$-stable process, one further applies the rejection method of \cite{Rosinski} to obtain (approximate) samples from the time  change process of the CGMY model. \cite{BallottaKyriakou} have developed a sampling method based on inverse Fourier transformation and the computing technique of fast Fourier transformation (FFT). This method involves three layers of approximation errors, namely, the regularization error from approximating the distribution of the CGMY (or the CGMY time change) increment by a regularization technique (cf. \cite{Hughett}), the truncation error from truncating the infinite integration domain of the inverse Fourier transformation integral, and the discretization error from applying the FFT technique. In simulation pricing of derivatives, it is difficult to quantify and bound the biases of Monte Carlo price estimates caused by the aforementioned specific approximation errors over the whole parameter space. \cite{PoirotTankov} developed an exact simulation pricing method, which does not introduce biases in price estimates, by exploiting the fact that under an appropriate change of measure, a CGMY process is a stable process whose increments can be sampled exactly; however, their method does not provide direct access to the sample paths of a CGMY process.


In pricing such path-dependent options as lookback and barrier options, \cite{KimKim2016} have recently developed a bridge sampling scheme (although only for the finite-variation case) that can lead to savings of simulation costs when combined with adaptive sampling techniques and to variance reduction when combined with stratified sampling techniques. This bridge sampling method is based upon saddle-point approximations for the related probability density functions and is otherwise comparable in costs and accuracy to the existing rejection sampling method when generating a fixed number of observations. However,  extension of this bridging sampling scheme to the infinite-variation case is nontrivial and has yet to be done.

In this paper, we develop a new and easy-to-implement sequential sampling method for  CGMY models   with either finite or infinite variation. As we shall see, our method involves only one simple error term, which has a transparent interpretation. To be specific, based on the time change representation of a CGMY process presented in \cite{MadanYor}, we find that the time change subordinator can be further decomposed into two independent components, namely, a finite generalized gamma convolution\footnotemark\footnotetext{The generalized gamma convolution law was introduced by \cite{Thorin} and studied by \cite{Bondesson}. See also, e.g., pp. 351--354 in \cite{JRY} for a comprehensive review on this class of distributions and processes.}  subordinator and an error term. For the first component, the increment of a finite generalized gamma convolution subordinator can be represented in distribution  as the product of a gamma random variable and another independent Dirichlet mean random variable (see, e.g., \cite{JamesBernoulli,JamesLamperti}). While the gamma random variable can be generated by standard procedures, the Dirichlet mean random variable can be sampled exactly via the double CFTP scheme of \cite{DevroyeJames}. As far as the error term is concerned, we show that it can be bounded and made arbitrarily small in  the $L^1$ sense.

In simulating Dirichlet mean random variables, the (exact) double CFTP method may have an excessive computational budget for certain parameter ranges. To reduce simulation costs with virtually no  loss of accuracy, an approximation scheme can be adopted instead of the double CFTP scheme. This approximate sampling method utilizes a special series representation of the Dirichlet mean random variable that converges exponentially fast, allowing approximation errors to be easily kept arbitrarily small.

We close this section by summarizing the following aspects of our contribution:
\begin{itemize}
\item The contribution of this paper lies more on the theoretical side than on the computational side. We have discovered a new path simulation method for CGMY processes with either finite or infinite variation. The method is built on a novel probabilistic result on the decomposition of the CGMY time change.
\item The method enjoys the unique feature that the upper bounds of the involved specific errors in different steps admit \emph{closed-form} expressions as functions of both the model and error parameters (see (\ref{eq:errorbound}), (\ref{eq:AppErrorBound}), (\ref{eq:L2boundCGMYAppError}) and Section \ref{subsec:error}), and more importantly, are \emph{explicitly} related to
    the bound of the simulation bias\footnotemark\footnotetext{We use ``simulation bias'' to refer to the error that one ultimately wants to control. It differs from other specific errors involved in different steps of a method. Taking simulation-based mean estimation for example, one naturally cares about the bias which is given by the difference between the true population mean and the mean of the approximate variable. This bias is controlled by the $L^1$ distance between the target and approximate random variables.}, which is measured by, e.g., the $L^1$ distance between the approximate variable and the target CGMY increment (see Section \ref{subsec:TransInterpError} and the discussion immediately following (\ref{eq:DirichletAppErrBound})).

\item However, existing methods, which must entail approximations in the infinite-variation case, do not share the above unique feature of our method. That is, for these methods, it is \emph{unclear} how the bound of the simulation bias is explicitly related to the bounds of various specific errors, which, moreover, do not admit closed-form expressions.
\item The above unique feature of our method is important. On the one hand, closed-form expressions of error bounds lead to closed-form solutions of optimal error parameters given pre-specified error tolerance levels without resort to further numerical procedures (cf., e.g., Section 4.2 of \cite{BallottaKyriakou}) that cost
    additional computational budget. On the other hand, the lack of an explicit relation between the bound of the simulation bias and the bounds of various specific errors could lead to setting either over optimistic (large) error tolerance levels that yield large biases or over conservative (small) error tolerance levels that yield extra computing costs. See Section \ref{sec:numerical} for an illustration of this point via simulations.
\item 
    The main message on the computational side from this paper is that knowing explicit relation between the simulation biases and the bounds of various specific errors is not less important than pursuing computing efficiency. We simply adopt existing algorithms for simulating the finite generalized gamma convolution component of the time change. Further reducing computational complexity of these algorithms is of course of great practical interest in its own right and warrant future study. Nonetheless, as can be seen from the simulation results in Section \ref{subsec:simresults}, under the studied model, our method with approximation scheme is advantageous in terms of computing speed over the two methods under comparison in achieving a same level of estimation accuracy.
\end{itemize}

The rest of this paper is organized as follows. In Section \ref{CGMYintro}, we provide a brief introduction to the CGMY model and related derivatives pricing problems. Section \ref{subsec:exactmethodYzeroone} provides an exact path simulation method for the finite-variation case, which is less familiar in the context of CGMY processes than in the context of tempered stable processes. The main results on the decomposition of the CGMY time change and their proofs are given in Section \ref{subsec:between02}. Section \ref{sec:TwoSimMethods} provides two schemes for simulating the finite generalized gamma convolution component of the time change. Section \ref{sec:numerical} is devoted to numerical studies where we compare our method with existing methods. We conclude in Section \ref{sec:conclusion}. The sampling algorithms are presented in the appendices.

\section{THE CGMY OPTION VALUATION MODEL}\label{CGMYintro}
A CGMY process $X=\{X(t),t\geq0\}$ is a pure jump L\'evy process with $X(0)=0$ and L\'evy density
\begin{align}\label{eq:LevyDensityCGMY}
\nu_X(x)=C\left(\frac{e^{-G|x|}}{|x|^{1+Y}}I\left\{x<0\right\}+\frac{e^{-Mx}}{x^{1+Y}}I\left\{x>0\right\}\right),
\end{align}
where $C>0,$ $G\geq0,$ $M\geq0,$ and $0<Y<2$ are four parameters. $Y$ is usually referred to as the \emph{stability index}. When $0<Y<1$ (respectively, $1\leq Y<2$), the CGMY process is of finite (respectively infinite) variation.

The characteristic function ($Y\neq1$) of $X(t)$ is given by
\begin{align}\label{eq:CharacteristicFunctionCGMY}
{\rm E}\left[e^{iuX(t)}\right]=\exp\left\{tC\Gamma(-Y)\left[(G+iu)^Y-G^Y+(M-iu)^Y-M^Y\right]\right\}.
\end{align}
The risk-neutral asset price process $S=\{S(t),t\geq0\}$ under the CGMY model is defined as
\begin{align}\label{eq:RiskNeutralAssetPrice}
S(t):=S(0)\exp\{(\omega+r-q)t+X(t)\},
\end{align}
where $r$ is the (constant) risk-free interest rate, $q$ is the asset's continuously compounded dividend yield, and $\omega$ is chosen such that the discounted asset price is a martingale, or, in other words,
$$
{\rm E}\left[\exp(\omega t+X(t))\right]=1.
$$
This condition and (\ref{eq:CharacteristicFunctionCGMY}) imply that
\begin{align}
\omega=-C\Gamma(-Y)\left[(G+1)^Y-G^Y+(M-1)^Y-M^Y\right],\notag
\end{align}
where $M\geq1$ is required to ensure that ${\rm E}[S(t)]<\infty$ for all $t\geq0$.

The present fair value $c$ of a derivative contract with general payoff $f(\{S(t),0\leq t\leq T\})$ at maturity $T$ is given by
$$
c={\rm E}[e^{-rT}f(\{S(t),0\leq t\leq T\})].
$$
Different forms of payoff function $f(\cdot)$ correspond to different derivatives contracts. If one can perfectly generate, say, $I$ independent and identically distributed (i.i.d.) sample paths $(S^{(i)})_{1\leq i\leq I}$ from $S$, then the Monte Carlo estimate of the derivative price $c$ is given as follows:
\begin{align}\label{eq:mcmcprice}
\hat{c}=\frac{1}{I}\sum_{i=1}^Ie^{-rT}f(\{S^{(i)}(t),0\leq t\leq T\}).
\end{align}

Let $0\equiv t_0<t_1<t_2<\ldots<t_n\equiv T$ be discrete monitoring times, $K$ the strike price, and $\mathcal{B}$  a prescribed barrier level; then the following provide four examples of payoff functions for four different derivatives contracts:
\begin{itemize}
\item European plain vanilla call option: $f(\{S(t),0\leq t\leq T\})=(S(T)-K)^+$;
\item floating strike lookback call option: $f(\{S(t),0\leq t\leq T\})=(S(T)-\max_{0\leq i\leq n}S(t_i))^+$;
\item up-and-in call barrier option: $f(\{S(t),0\leq t\leq T\})=(S(T)-K)^+1_{\{\max_{0\leq i\leq n}S(t_i)>\mathcal{B}\}}$;
\item Asian call option with discrete monitoring: $f(\{S(t),0\leq t\leq T\})=\left(\frac{1}{n+1}\sum_{i=0}^nS(t_i)-K\right)^+$.
\end{itemize}
In all cases, Monte Carlo simulation pricing reduces to simulation of increments of the log return process $X$ in (\ref{eq:RiskNeutralAssetPrice}). In the following sections, we introduce new sequential sampling schemes for simulating increments of the CGMY log return process.

\section{AN EXACT METHOD FOR $Y\in(0,1)$}\label{subsec:exactmethodYzeroone}
It is notable that exact simulation schemes are less familiar in the context of CGMY processes than in the context of tempered stable processes. In this section, we elaborate on how exact sampling methods for tempered stable processes can be adapted to sampling CGMY increments in the finite-variation case.

From (\ref{eq:LevyDensityCGMY}), $X$ has the following difference-of-CGMY  representation:
\begin{align}
X(t)=X^+(t)-X^-(t),\notag
\end{align}
where $X^+=\{X^+(t),t\geq0\}$ and $X^-=\{X^-(t),t\geq0\}$ are two independent L\'evy processes with L\'evy densities
\begin{align}\label{eq:differenceCGMYlevyDens}
\nu_{X^+}(x)=C\frac{e^{-Mx}}{x^{1+Y}}I\left\{x>0\right\} ~~~{\rm and~~~}\nu_{X^-}(x)=C\frac{e^{-Gx}}{x^{1+Y}}I\left\{x>0\right\},
\end{align}
respectively.

When $0<Y<1$, the exact sampling method of \cite{Devroye} for exponentially tilted stable distributions can be adapted to the simulation for the CGMY increments. Observe that the CGMY process with $0<Y<1$ can be represented as a difference of two independent increasing L\'evy processes with L\'evy densities (\ref{eq:differenceCGMYlevyDens}). Hence, it suffices to consider the simulation problem for these increasing positive processes, i.e., subordinators. We take the simulation of process $X^+$ with L\'evy density $\nu_{X^+}(x)$ for illustration. Then the simulation of process $X^-$ with L\'evy density $\nu_{X^-}(x)$ follows similarly. As a consequence, the CGMY process $X$ is given by the difference between $X^+$ and $X^-$.

Without loss of generality, we only need  consider the simulation of variables $X^+(t)$ for $t>0$, because L\'evy processes have the stationary increments property. The distribution of $X^+(t)$ is exponentially tilted stable, i.e.,
$$
X^+(t)\overset{d}=\lambda^{1/Y}\mathcal{S}_{Y,M\lambda^{1/Y}},
$$
where $\lambda:=tC\Gamma(1-Y)/Y$, and $\mathcal{S}_{Y,M\lambda^{1/Y}}$ is an exponentially tilted stable random variable that has density function $$e^{M^Y\lambda-M\lambda^{1/Y}x}g_{Y}(x),~~~ x>0,$$
where $g_{Y}(x)$ is the density of the unilateral $Y$-stable random variable with Laplace transform
$$\int_0^\infty e^{-\mu x}g_Y(x)\,dx=e^{-\mu^Y},~~~ \mu>0.$$

\cite{Devroye} proposed a double rejection method for the exact simulation of an exponentially tilted stable random variable that can be uniformly fast over all parameter ranges. Based on \cite{Devroye},  a double rejection algorithm for generating random variable $X^+(t)$ is given in Section \ref{subsec:appdoublerejection}.

\section{THE MAIN RESULTS FOR $Y\in(0,2)$}\label{subsec:between02}
We now turn to the general case where the CGMY process can be of infinite variation.
\subsection{A Decomposition of the CGMY Time Change}
\cite{MadanYor} showed that a CGMY process can be represented as a time-changed Brownian motion as follows:
\begin{align}\label{eq:TimeChangedCGMY}
X(t)=\theta \mathcal{T}(t)+W(\mathcal{T}(t)),
\end{align}
where $\theta=(G-M)/2$, and $W=\{W(t),t\geq0\}$ is a standard Brownian motion that is independent of the time change subordinator $\mathcal{T}=\{\mathcal{T}(t),t\geq0\}.$ \cite{MadanYor} identified the L\'evy density of $\mathcal{T}$ as follows:
\begin{align}\label{eq:LevyDensTimeChangeCGMY}
\nu_{\mathcal{T}}(x)=C\frac{2^{Y/2-1}\Gamma(Y/2)}{\Gamma(Y)}\frac{e^{-(\tilde{\theta}^2-\theta^2)x/2}}{x^{1+Y/2}}{\rm E}\left[\exp\left(-\frac{\tilde{\theta}^2x}{2}\frac{\gamma_{Y/2}}{\gamma_{1/2}}\right)\right],
\end{align}
where $\tilde{\theta}=(G+M)/2$, and $\gamma_{Y/2}$ and $\gamma_{1/2}$ are independent gamma random variables with unit scales and shapes $Y/2$ and $1/2$, respectively.\\

The following theorem provides a decomposition of the time change $\mathcal{T}$, facilitating the path simulation of both finite- and infinite-variation CGMY processes.
\begin{theorem}\label{prop1}
For $L>0$ and fixed $t>0$, the time change subordinator $\mathcal{T}(t)$ in (\ref{eq:TimeChangedCGMY}) has the following decomposition:
\begin{align}
\mathcal{T}(t)=\mathcal{T}_L(t)+\epsilon_L(t),\label{eq:timechangedecomposition}
\end{align}
where $\mathcal{T}_L(t)$ and $\epsilon_L(t)$ are independent and have the following distributional properties:
\begin{itemize}
\item[(i)] $\mathcal{T}_L(t)$ is a generalized gamma convolution random variable that has Laplace exponent
    $$-\log\left({\rm E}\left[e^{-\mu\mathcal{T}_L(t)}\right]\right)=\frac{2\widetilde{C}L^{Y/2}}{Y}{\rm E}\left[\log\left(1+\mu R\right)\right],$$
    where $\mu>0$, $\widetilde{C}:=tC2^{Y/2-1}/\Gamma(Y)$, and the random variable $R$ is given by
    $$
    R:=\frac{1}{\frac{GM+\tilde{\theta}^2\mathcal{R}}{2}+\mathcal{Z}},
    $$
    where $\mathcal{R}:=\gamma_{Y/2}/\gamma_{1/2}$ is independent of $\mathcal{Z}$, which has probability density function $f_{\mathcal{Z}}(z)=YL^{-Y/2}/2z^{Y/2-1},$ $0\leq z\leq L$;
\item[(ii)] The standardized $\epsilon_L(t)$ has a standard normal limiting distribution as $L\to\infty$, i.e.,
    $$
\frac{\epsilon_L(t)-{\rm E}\left(\epsilon_L(t)\right)}{\sqrt{{\rm Var}\left(\epsilon_L(t)\right)}}\overset{d}\to N(0,1).
$$
 In particular,
\begin{align}{\rm E}[\epsilon_L(t)^2]\leq\frac{\widetilde{C}^2}{(1-Y/2)^2L^{2-Y}}+\frac{\widetilde{C}}{(2-Y/2)L^{2-Y/2}}.\label{eq:errorbound}\end{align}
\end{itemize}
\end{theorem}

From (\ref{eq:errorbound}), we can see that $\epsilon_L(t)=O_p(1/L^{1-Y/2})$ and $L$ is an \emph{error parameter} that controls the magnitude of the error. Hence, we can choose some large $L$ such that $\epsilon_L(t)$ is negligible. Then we can use samples of $\mathcal{T}_L(t)$ to approximate that of $\mathcal{T}(t)$ given that we can simulate $\mathcal{T}_L(t)$ perfectly. In Section \ref{subsec:perfectsimGGC}, we show that exact simulation of $\mathcal{T}_L(t)$ is possible. Before introducing the simulation methods, we first provide a proof of Theorem \ref{prop1} in the next section.

\subsection{Proof of Theorem \ref{prop1}}\label{App:appendix}
Recall that in Theorem \ref{prop1}, for notational clarity, we let $$\widetilde{C}:=tC2^{Y/2-1}/\Gamma(Y)~~~{\rm and~~~}\mathcal{R}:=\gamma_{Y/2}/\gamma_{1/2},$$
where $\gamma_{Y/2}$ and $\gamma_{1/2}$ are independent gamma random variables as given in (\ref{eq:LevyDensTimeChangeCGMY}).

For $\mu>0$, let $\varphi_{\mathcal{T}(t)}(\mu):=-\log({\rm E}[e^{-\mu\mathcal{T}(t)}])$ denote the Laplace exponent of $\mathcal{T}(t)$. We have
\begin{align}
\varphi_{\mathcal{T}(t)}(\mu)&=tC\frac{2^{Y/2-1}\Gamma(Y/2)}{\Gamma(Y)}\int_0^\infty(1-e^{-\mu x})\frac{e^{-(\tilde{\theta}^2-\theta^2)x/2}}{x^{1+Y/2}}{\rm E}\left[e^{-\tilde{\theta}^2\mathcal{R}x/2}\right]dx\notag\\
&=tC\frac{2^{Y/2-1}\Gamma(Y/2)}{\Gamma(Y)}{\rm E}\left[\int_0^\infty(1-e^{-\mu x})\frac{e^{-(\tilde{\theta}^2-\theta^2+\tilde{\theta}^2\mathcal{R})x/2}}{x^{1+Y/2}}\,dx\right]\notag\\
&=\widetilde{C}{\rm E}\left[\int_0^\infty(1-e^{-\mu x})\frac{1}{x}\int_0^\infty e^{-\left(\frac{\tilde{\theta}^2-\theta^2+\tilde{\theta}^2\mathcal{R}}{2}+z\right)x}z^{Y/2-1}\,dz\,dx\right]\notag\\
&=\widetilde{C}\int_0^\infty{\rm E}\left[\log\left(1+\frac{\mu}{\frac{\tilde{\theta}^2-\theta^2+\tilde{\theta}^2\mathcal{R}}{2}+z}\right)\right]\frac{1}{z^{1-Y/2}}\,dz\notag.
\end{align}

For $L>0$, $\varphi_{\mathcal{T}(t)}(\mu)$ can be written as follows:
\begin{align}
\varphi_{\mathcal{T}(t)}(\mu)&=\underbrace{\widetilde{C}\int_0^L{\rm E}\left[\log\left(1+\frac{\mu}{\frac{\tilde{\theta}^2-\theta^2+\tilde{\theta}^2\mathcal{R}}{2}+z}\right)\right]\frac{1}{z^{1-Y/2}}\,dz}_{\varphi_{L,1}(\mu)}\notag\\
&~~~+\underbrace{\widetilde{C}\int_L^\infty{\rm E}\left[\log\left(1+\frac{\mu}{\frac{\tilde{\theta}^2-\theta^2+\tilde{\theta}^2\mathcal{R}}{2}+z}\right)\right]\frac{1}{z^{1-Y/2}}\,dz}_{\varphi_{L,2}(\mu)}\notag.
\end{align}
This implies that $\mathcal{T}(t)$ can be decomposed as follows:
$$
\mathcal{T}(t)=\mathcal{T}_L(t)+\epsilon_L(t),
$$
where $\mathcal{T}_L(t)$ is independent of $\epsilon_L(t)$. $\varphi_{L,1}(\mu)$ and $\varphi_{L,2}(\mu)$ are the Laplace exponents of $\mathcal{T}_L(t)$ and $\epsilon_L(t)$, respectively.\\

First, with some algebra, $\varphi_{L,1}(\mu)$ can be rewritten as
\begin{align}
\varphi_{L,1}(\mu)&=\frac{2\widetilde{C}L^{Y/2}}{Y}{\rm E}\left\{{\rm E}\left[\left.\log\left(1+\frac{\mu}{\frac{\tilde{\theta}^2-\theta^2+\tilde{\theta}^2\mathcal{R}}{2}+\mathcal{Z}}\right)\right|\mathcal{Z}\right]\right\}\notag\\
&=\frac{2\widetilde{C}L^{Y/2}}{Y}{\rm E}\left[\log\left(1+\mu R\right)\right]\notag,
\end{align}
where $\mathcal{Z}$ is independent of $\mathcal{R}$ with probability density function $f_{\mathcal{Z}}(z)=YL^{-Y/2}/2z^{Y/2-1},$ $0\leq z\leq L$, and
$$
R=\frac{1}{\frac{\tilde{\theta}^2-\theta^2+\tilde{\theta}^2\mathcal{R}}{2}+\mathcal{Z}}=\frac{1}{\frac{GM+\tilde{\theta}^2\mathcal{R}}{2}+\mathcal{Z}},
$$
where the second equality follows from the fact that $\tilde{\theta}^2-\theta^2=GM$. The support of the distribution of $R$ is $[0,2/(GM)]$, and hence the random variable $R$ is bounded by $2/(GM)$. By, for example, eq. (25) on p. 354 of \cite{JRY}, the random variable $\mathcal{T}_L(t)$ with Laplace exponent $\varphi_{L,1}(\mu)$ is a generalized gamma convolution random variable.\\

Second, define
\begin{align}\label{eq:DefMuLSigL}
\mu_L:=\widetilde{C}{\rm E}\left(\int_L^\infty\frac{1}{\frac{GM+\tilde{\theta}^2\mathcal{R}}{2}+z}\frac{1}{z^{1-Y/2}}\,dz\right)~~~{\rm and}~~~\sigma^2_L:=\widetilde{C}{\rm E}\left(\int_L^\infty\frac{1}{\left(\frac{GM+\tilde{\theta}^2\mathcal{R}}{2}+z\right)^2} \frac{1}{z^{1-Y/2}}\,dz\right).
\end{align}
$\mu_L$ and $\sigma^2_L$ are the mean and variance of $\epsilon_L(t)$. The exact evaluation of $\mu_L$ and $\sigma^2_L$ is difficult. However, we can easily find their upper bounds:
$$
\mu_L\leq\widetilde{C}\int_L^\infty z^{Y/2-2}dz=\frac{\widetilde{C}}{(1-Y/2)L^{1-Y/2}},
$$
and, similarly,
$$
\widetilde{C}{\rm E}\left [\frac{\left(\frac{GM+\tilde{\theta}^2\mathcal{R}}{2}+L\right)^{Y/2-2}}{2-Y/2}\right]\leq\sigma^2_L\leq\frac{\widetilde{C}}{(2-Y/2)L^{2-Y/2}}.
$$
Hence, we have $L\sigma_L\to\infty$ as $L\to\infty$.

Because $L\sigma_L\to\infty$ as $L\to\infty$, the random variable $\epsilon_L(t)$ with Laplace exponent $\varphi_{L,2}(\mu)$ can be approximated by a normal random variable with mean $\mu_L$ and variance $\sigma^2_L$. To see this, recall that
\begin{align}
\varphi_{L,2}(\mu)&=\widetilde{C}\int_L^\infty{\rm E}\left[\log\left(1+\frac{\mu}{\frac{\tilde{\theta}^2-\theta^2+\tilde{\theta}^2\mathcal{R}}{2}+z}\right)\right]\frac{1}{z^{1-Y/2}}\,dz\notag.
\end{align}
Therefore, the Laplace exponent of the standardized $\epsilon_L(t)$, i.e., $(\epsilon_L(t)-\mu_L)/\sigma_L$, is given by
\begin{align}\label{eq:CLTerror}
-\mu\frac{\mu_L}{\sigma_L}+\varphi_{L,2}(\mu/\sigma_L)&=-\mu\frac{\mu_L}{\sigma_L}+\widetilde{C}\int_L^\infty{\rm E}\left[\log\left(1+\frac{\mu}{\left(\frac{\tilde{\theta}^2-\theta^2+\tilde{\theta}^2\mathcal{R}}{2}+z\right)\sigma_L}\right)\right]\frac{1}{z^{1-Y/2}}\,dz.
\end{align}
By Taylor's theorem with mean-value form of the remainder and again by the fact that $\tilde{\theta}^2-\theta^2=GM$, the second term on the right hand side of equation (\ref{eq:CLTerror}) can be written as follows
\begin{align}
&~~~\widetilde{C}\int_L^\infty{\rm E}\left[\log\left(1+\frac{\mu}{\left(\frac{GM+\tilde{\theta}^2\mathcal{R}}{2}+z\right)\sigma_L}\right)\right]\frac{1}{z^{1-Y/2}}\,dz\notag\\
&=\widetilde{C}\int_L^\infty{\rm E}\left[\frac{\mu}{\left(\frac{GM+\tilde{\theta}^2\mathcal{R}}{2}+z\right)\sigma_L}-\frac{1}{2}\frac{\mu^2}{\left(\frac{GM+\tilde{\theta}^2\mathcal{R}}{2}+z\right)^2\sigma_L^2}\right]\frac{1}{z^{1-Y/2}}\,dz\notag\\
&~~~+\widetilde{C}\int_L^\infty{\rm E}\left[\frac{1}{3(1+x^*)^3}\frac{\mu^3}{\left(\frac{GM+\tilde{\theta}^2\mathcal{R}}{2}+z\right)^3\sigma_L^3}\right]\frac{1}{z^{1-Y/2}}\,dz\notag\\
&=\frac{\mu}{\sigma_L}\widetilde{C}{\rm E}\left(\int_L^\infty\frac{1}{\frac{GM+\tilde{\theta}^2\mathcal{R}}{2}+z}\frac{1}{z^{1-Y/2}}\,dz\right)-\frac{\mu^2}{2\sigma_L^2}\widetilde{C}{\rm E}\left(\int_L^\infty\frac{1}{\left(\frac{GM+\tilde{\theta}^2\mathcal{R}}{2}+z\right)^2} \frac{1}{z^{1-Y/2}}\,dz\right)\notag\\
&~~~+\underbrace{\widetilde{C}\int_L^\infty{\rm E}\left[\frac{1}{3(1+x^*)^3}\frac{\mu^3}{\left(\frac{GM+\tilde{\theta}^2\mathcal{R}}{2}+z\right)^3\sigma_L^3}\right]\frac{1}{z^{1-Y/2}}\,dz}_{{\rm remainder}}\notag\\
&=\frac{\mu\mu_L}{\sigma_L}-\frac{\mu}{2}+O\left(\frac{1}{L\sigma_L}\right),\label{eq:CLTlaplaceexponentcalu}
\end{align}
where $x^*$ is some variable between 0 and $\mu/(((GM+\tilde{\theta}^2\mathcal{R})/2+z)\sigma_L)$, the last equality follows from the definitions in (\ref{eq:DefMuLSigL}) and the following approximation of the remainder term
\begin{align}
&~~~\widetilde{C}\int_L^\infty{\rm E}\left[\frac{1}{3(1+x^*)^3}\frac{\mu^3}{\left(\frac{GM+\tilde{\theta}^2\mathcal{R}}{2}+z\right)^3\sigma_L^3}\right]\frac{1}{z^{1-Y/2}}\,dz\notag\\
&\leq\frac{\mu^3}{3\sigma_L^2}\widetilde{C}{\rm E}\left(\int_L^\infty\frac{1}{\left(\frac{GM+\tilde{\theta}^2\mathcal{R}}{2}+z\right)^2} \frac{1}{z^{1-Y/2}}\,dz\right)\times\frac{1}{L\sigma_L}=O\left(\frac{1}{L\sigma_L}\right)\notag.
\end{align}
Substituting (\ref{eq:CLTlaplaceexponentcalu}) for the last term in (\ref{eq:CLTerror}) leads to the following Laplace exponent of $(\epsilon_L(t)-\mu_L)/\sigma_L$,
$$
-\frac{\mu}{2}+O\left(\frac{1}{L\sigma_L}\right),
$$
which converges to $-\mu/2$ as $L\to\infty$, since $L\sigma_L\to\infty$. We have thus proved
$$
\frac{\epsilon_L(t)-{\rm E}\left(\epsilon_L(t)\right)}{\sqrt{{\rm Var}\left(\epsilon_L(t)\right)}}\overset{d}\to N(0,1),
$$
completing the proof of Theorem 1.

\subsection{The Error Term $\epsilon_L(t)$}\label{subsec:error}
In this section, we study the error term $\epsilon_L(t)$  in the decomposition (\ref{eq:timechangedecomposition}) of Theorem \ref{prop1}. For convenience, we recall here the inequality (\ref{eq:errorbound}):
$$
{\rm E}[\epsilon_L(t)^2]\leq\frac{\widetilde{C}^2}{(1-Y/2)^2L^{2-Y}}+\frac{\widetilde{C}}{(2-Y/2)L^{2-Y/2}},
$$
where $\widetilde{C}:=tC2^{Y/2-1}/\Gamma(Y)$. This inequality provides an upper bound on the second moment of $\epsilon_L(t)$. Notice that this error bound admits a closed-form expression as a function of both the error parameter $L$ and model parameters. Holding the model parameters and $t$ constant, for any pre-specified small error tolerance level $\varepsilon>0$, one can choose $L$ such that both two terms on the right-hand side of (\ref{eq:errorbound}) are less than or equal to $\varepsilon^2/2$. The smallest (optimal) $L$ that satisfies this requirement is given by
\begin{align}\label{eq:Lchoice}
L_{\rm min}=\max\left\{\left(\frac{2\widetilde{C}^2}{\varepsilon^2(1-Y/2)^2}\right)^{1/(2-Y)},\left(\frac{2\widetilde{C}}{\varepsilon^2(2-Y/2)}\right)^{1/(2-Y/2)}\right\}.
\end{align}
If $L$ is chosen as in (\ref{eq:Lchoice}), then, by Jensen's inequality, we have
$$
{\rm E}|\epsilon_L(t)|\leq\varepsilon.
$$
Hence, for $\epsilon_L(t)$ to be negligible, one only needs prescribe an error tolerance level $\varepsilon$ and then choose $L=L_{\rm min}$ as above. Of course, smaller $\varepsilon$ leads to larger $L_{\rm min}$ and hence, as we shall see in Section \ref{subsec:perfectsimGGC}, greater computational effort in simulating $\mathcal{T}_L(t)$ using the double CFTP scheme. Moreover, when $\varepsilon$, $t$, and other model parameters are fixed, $L_{\rm min}$ increases with $Y$, and hence, in order to achieve a same precision, the double CFTP method is more time-consuming for larger $Y$ than for smaller $Y$. The situation can be challenging when $Y$ approaches 2.

\subsection{Bound The Error of Approximating CGMY Increment Explicitly}\label{subsec:TransInterpError}
Recall that the CGMY increment has the time change representation (\ref{eq:TimeChangedCGMY}), i.e., $X(t)=\theta \mathcal{T}(t)+W(\mathcal{T}(t))$, and, from Theorem \ref{prop1}, the time change has the decomposition $\mathcal{T}(t)=\mathcal{T}_L(t)+\epsilon_L(t)$. We can thus write the CGMY increment as follows:
 $$X(t)\overset{d}=\underbrace{\theta\mathcal{T}_L(t)+\sqrt{\mathcal{T}_L(t)}W_1(1)}_{X_1(t)}+\underbrace{\theta\epsilon_L(t)+\sqrt{\epsilon_L(t)}W_2(1)}_{X_2(t)},$$
where $W_1(1)$ and $W_2(1)$ are two independent standard normal random variables that are independent of the remaining random variables on the right-hand side of the above equation. Therefore, sampling from $X(t)$ is equivalent to sampling from the sum of two independent variables $X_1(t)$ and $X_2(t)$. Based on this observation, we suggest approximate the distribution of $X(t)$ by that of $X_1(t)$ from which one can perfectly simulate since one can perfectly sample from the distribution of $\mathcal{T}_L(t)$ as we shall see in Section \ref{subsec:perfectsimGGC}. Then $X_2(t)$ can be deemed as the error (or residual) of approximating $X(t)$ by $X_1(t)$. By simple calculation, the $L^1$ mean of this error (i.e., the $L^1$ distance between the approximate variable $X_1(t)$ and the target CGMY increment $X(t)$) is given as follows
\begin{align}\label{eq:L2boundCGMYAppError}
{\rm E}\left(|X_2(t)|\right)\leq|\theta|{\rm E}(\epsilon_L(t))+\sqrt{{\rm E}\left(\epsilon_L(t)\right)}\leq|\theta|\varepsilon+\sqrt{\varepsilon},
\end{align}
provided that $L$ is chosen as in (\ref{eq:Lchoice}).
The inequality in (\ref{eq:L2boundCGMYAppError}) shows that the upper bound of the $L^1$ distance between the approximate variable $X_1(t)$ and the target CGMY increment $X(t)$ can be \emph{explicitly} expressed in closed-form as a function of the pre-specified tolerance level $\varepsilon$ for the error involved in simulating the time change. For existing methods that entail approximations, there are no such closed-form relations which are key in determining the right (optimal) choices of tolerance levels (or equivalently, error parameters) for specific errors involved in different steps of a simulation method to avoid either large simulation biases or extra computing costs. In this sense, the errors involved in our method have a more transparent interpretation (see also the discussion immediately following (\ref{eq:DirichletAppErrBound})) than existing methods.


\section{Simulation of $\mathcal{T}_L(t)$}\label{sec:TwoSimMethods}
In this section, we introduce two methods for simulating the finite generalized gamma convolution component of the time change, i.e., $\mathcal{T}_L(t)$. One method is exact and the other one is approximate. We show that the approximation scheme is accurate and can be faster than the exact method. These existing sampling algorithms are by no means optimal in terms of computational complexity, further research should be done in reducing computing costs. However, this is not straightforward and beyond the scope of this paper which focuses on the theoretical probabilistic results.

\subsection{Perfect Simulation}\label{subsec:perfectsimGGC}
We first explain how $\mathcal{T}_L(t)$ can be exactly sampled. Let $$\tau:=\frac{2\widetilde{C}L^{Y/2}}{Y}.$$
We have shown in Theorem \ref{prop1} that $\mathcal{T}_L(t)$ is a generalized gamma convolution random variable with Laplace exponent
\begin{align}
\tau{\rm E}\left[\log\left(1+\mu R\right)\right].\label{eq:laplaceexpGGC}
\end{align}
From \cite{JamesBernoulli}, a generalized gamma convolution random variable $\mathcal{T}_L(t)$ with Laplace exponent (\ref{eq:laplaceexpGGC}) has the following representation (see also \cite{JRY}):
\begin{align}\label{eq:gammamitureofrepGGC}
\mathcal{T}_L(t)\overset{d}=\gamma_{\tau}\cdot \mathcal{D}_{\tau}(F_R),
\end{align}
where $\gamma_{\tau}$ is independent of $\mathcal{D}_{\tau}(F_R)$, $\gamma_{\tau}$ is a gamma random variable with shape $\tau$ and unit scale, and $\mathcal{D}_{\tau}(F_R)$ ($F_R$ denotes the cumulative distribution function of random variable $R$) is a Dirichlet mean random variable that solves for random variable $\mathcal{D}$ in the following stochastic equation (see \cite{JamesLamperti}):
\begin{align}\label{eq:dmeanstochasticequation}
\mathcal{D}\overset{d}=\beta_{1,\tau}R+(1-\beta_{1,\tau})\mathcal{D},
\end{align}
where $\beta_{1,\tau}$ is a beta random variable with parameter values ($1,\tau$), and the random variables on the right-hand side of  (\ref{eq:dmeanstochasticequation}) are independent of one another.

By (\ref{eq:gammamitureofrepGGC}), simulation of $\mathcal{T}_L(t)$ reduces to  simulation of a gamma random variable $\gamma_{\tau}$, which is available in most standard numerical libraries, and  simulation of a Dirichlet mean random variable $\mathcal{D}_{\tau}(F_R)$, which we elaborate upon below.

\cite{DevroyeJames} devised an exact sampler termed  double CFTP (``coupling from the past'') for generating random numbers from the steady-state Markov chain distribution (of $\mathcal{D}$) determined by the following generic stochastic equation:
\begin{align}\label{eq:srochasticequationgeneral}
\mathcal{D}\overset{d}=BQ+(1-B)\mathcal{D},
\end{align}
where  double CFTP requires that the density function $h(\cdot)$ of the random variable $B$ can be precisely evaluated and is bounded from below on $[0,1]$ by a constant $c_h>0$, $0<Q\leq c_Q<\infty$, with $c_Q$ being a constant, and again the random variables on the right-hand side of the above equation are independent of one another. The double CFTP algorithm for generating random numbers from $\mathcal{D}$ in (\ref{eq:srochasticequationgeneral}) is given in Section \ref{subsec:doublecftp}.

When $B=\beta_{1,\tau}$ and $Q=R$ in (\ref{eq:srochasticequationgeneral}), we recover (\ref{eq:dmeanstochasticequation}), and the solution to $\mathcal{D}$ is just the Dirichlet mean random variable $\mathcal{D}_{\tau}(F_R)$.  Because the density function of $\beta_{1,\tau}$ takes the form $h(x)=\tau(1-x)^{\tau-1}$ for $x\in[0,1]$ and $R\leq2/(GM)$, in the Dirichlet mean case, the requirements of the double CFTP scheme, namely, that the density function $h(\cdot)$ of the random variable $B$ can be precisely evaluated and is bounded from below on $[0,1]$ by a constant $c_h>0$ and that $0<Q\leq c_Q<\infty$, are satisfied with $c_Q=2/(GM)$ and $c_h=\tau$ when $0<\tau\leq1$.

It appears that in practice, $0<\tau\leq1$ is a tight constraint for the double CFTP scheme to be applicable. Nonetheless, when $\tau>1$, we can always decompose $\mathcal{D}_{\tau}(F_R)$ as
$$
\mathcal{D}_{\tau}(F_R)\overset{d}=\sum_{j=1}^J\frac{\gamma_j\mathcal{D}_{\tau_j}(F_R)}{\gamma},
$$
where $J$ is an integer, $\tau=\sum_{j=1}^J\tau_j$ for $\tau_j>0$, $\gamma=\sum_{j=1}^J\gamma_j,$ $\gamma_j$ are independent gamma random variables with shapes $\tau_j$ and common unit scale, $\mathcal{D}_{\tau_j}(F_R)$ are independent Dirichlet mean random variables with shapes $\tau_j$ and common scale variable $R$, and $\gamma_j$ are independent of $\mathcal{D}_{\tau_j}(F_R)$. \cite{JKZ:2013} provide the optimal choices of $J$ and $\tau_j$ as $J=\lfloor \tau\rfloor+1$ and $\tau_j\equiv\tau/J$. Therefore, the requirement $0<\tau\leq1$ of  double CFTP poses no difficulty for the simulation of Dirichlet mean random variables with bounded scale random variable $R$.

Concern might be expressed about the effects of the range of $\tau$ on the computational complexity of simulations using the double CFTP sampler, since larger $\tau$ means that more random numbers need to be generated. Recall that $\tau=2\widetilde{C}L^{Y/2}/Y=(2L^{Y/2})tC2^{Y/2-1}/(\Gamma(Y)Y).$ For fixed $t$, $C$, and $L$, the denominator of $\tau$ , i.e., $\Gamma(Y)Y$, is bounded from below by a strictly positive constant for $Y\in[0,2]$, indicating that $\tau$ does not blow up with $Y$. Therefore, the computational complexity of a simulation depends mainly on $L$. To be precise, when $t$ and the  parameters $C$ and $Y$ are held constant, $\tau$ increases with $L$. $L$ is usually large, since this is necessary to ensure that the error term $\epsilon_L(t)$ in (\ref{eq:timechangedecomposition}) is negligible. It is easy to see that, with the remaining parameters held constant, $\tau$ increases faster with $L$ when $Y$ is larger.


\subsection{An Approximation Scheme}\label{subsubsec:APPscheme}
From the last paragraph of Section \ref{subsec:perfectsimGGC} and the discussions in Section \ref{subsec:error}, we notice that the suggested double CFTP scheme in Section \ref{subsec:perfectsimGGC} may be time-consuming for certain parameter ranges, for example, when $Y$ approaches 2 while  other parameters are held fixed. Therefore, we need to find an alternative method that can allow significant savings in simulation costs with virtually no loss of accuracy compared with the  exact simulation of $\mathcal{T}_L(t)$ provided by the double CFTP sampler. In this subsection, we shall introduce an approximation method that serves this purpose.

To understand the approximation scheme, we need to note that the Dirichlet mean random variable $\mathcal{D}_{\tau}(F_R)$ in (\ref{eq:gammamitureofrepGGC}) has the following series representation:
\begin{align}\label{eq:DmeanseriesRep}
\mathcal{D}_{\tau}(F_R)\overset{d}=\sum_{i=1}^\infty \tilde{B}_iR_i,
\end{align}
where
$$
\tilde{B}_1=B_1~~~{\rm and}~~~\tilde{B}_i=B_i\prod_{j=1}^{i-1}(1-B_j),~i\geq2,
$$
$B_i,~i=1,2,\ldots,$ are i.i.d. random variables equal in distribution to the beta random variable $\beta_{1,\tau}$ in (\ref{eq:dmeanstochasticequation}), and, independently, $R_i,~i=1,2,\ldots,$ are i.i.d. random variables that have the same distribution as the random variable $R$ defined in Theorem \ref{prop1}. The series representation (\ref{eq:DmeanseriesRep}) can be seen as a result of the definition of Dirichlet mean random variables in \cite{CifarelliRegazzini} and the stick-breaking random probability measures studied in \cite{IshwaranJames}, to which we refer for a complete history of those concepts.

Now we are ready to present the approximation scheme. Because $\sum_{i=1}^\infty\tilde{B}_i=1$ and the random variable $R$ is bounded by $2/(GM)$, the error induced by truncating (\ref{eq:DmeanseriesRep}) after, say, $n$ terms is bounded by $$2/(GM)\left(1-\sum_{i=1}^n\tilde{B}_i\right).$$ Hence, one solution to simulating $\mathcal{D}_{\tau}(F_R)$ is the stopping time approach of \cite{GHW}. To be specific, let
\begin{align}\label{eq:GHWstoppingN}
\mathcal{N}:=\min_n\left\{n:2/(GM)\left(1-\sum_{i=1}^n\tilde{B}_i\right)<\tilde{\varepsilon},~n=1,2,\ldots\right\},
\end{align}
which is a stopping time indicating when the tail of (\ref{eq:DmeanseriesRep}) falls below a small threshold (i.e., an error tolerance level) $\tilde{\varepsilon}$. The approximate variable $\mathcal{D}^{\mathcal{N}}_{\tau}(F_R)$ for $\mathcal{D}_{\tau}(F_R)$ is thus  given by
$$
\mathcal{D}^{\mathcal{N}}_{\tau}(F_R):=\sum_{i=1}^\mathcal{N} \tilde{B}_iR_i.
$$
The random number generation in sampling $\mathcal{D}^{\mathcal{N}}_{\tau}(F_R)$ is otherwise quite straightforward. The stopping rule (\ref{eq:GHWstoppingN}) leads to the following distance bound which is {\it exact} rather than in the $L^1$ sense:
\begin{align}\label{eq:DirichletAppErrBound}
|\mathcal{D}^{\mathcal{N}}_{\tau}(F_R)-\mathcal{D}_{\tau}(F_R)|\leq\tilde{\varepsilon}.
\end{align}
That is, a pre-specified error tolerance level $\tilde{\varepsilon}$ precisely gives an exact upper bound on the error of approximating $\mathcal{D}_{\tau}(F_R)$ by $\mathcal{D}^{\mathcal{N}}_{\tau}(F_R)$. Furthermore, from the decomposition of $\mathcal{T}_L(t)$ in (\ref{eq:gammamitureofrepGGC}), the above approximation introduces an additional error, which is similar to $X_2(t)$ in Section \ref{subsec:TransInterpError}, in simulating the CGMY increment $X(t)$. By similar arguments to that used in (\ref{eq:L2boundCGMYAppError}), this additional error can be bounded in the $L^1$ sense by
$$
|\theta|\tau\tilde{\varepsilon}+\sqrt{\tau\tilde{\varepsilon}}.
$$

The computational complexity of this method
depends on the upper bound on the tail of (\ref{eq:DmeanseriesRep}), i.e.,
$
2/(GM)(1-\sum_{i=1}^n\tilde{B}_i),
$
which together with $\tilde{\varepsilon}$ determine $\mathcal{N}$. By simple calculation, we find that the expectation of the upper bound on the tail of (\ref{eq:DmeanseriesRep}) is
\begin{align}\label{eq:AppErrorBound}
{\rm E}\left(2/(GM)\left(1-\sum_{i=1}^n\tilde{B}_i\right)\right)=\frac{2}{GM}\left(\frac{\tau}{1+\tau}\right)^n,
\end{align}
meaning that on average the tail of the series (\ref{eq:DmeanseriesRep}) decreases exponentially fast to zero (i.e., the series converges exponentially fast), provided that $2/(GM)$ and $\tau$ take moderately sized values.

\begin{remark}
When we consider the simulation of increments over small time intervals, which is especially pertinent to pricing (near) continuously monitored path-dependent options, $\tau$ usually takes moderately sized values. In this case, the approximation method described here is advantageous over the double CFTP method in terms of computational complexity (see, e.g., Sections 5.1 and 5.2 and, in particular, Remark 5.3 of \cite{JKZ:2013}).
\end{remark}

\section{SIMULATION STUDIES}\label{sec:numerical}
In this section, we show the importance of knowing closed-form relations between bounds of simulation biases and that of various specific errors (or tolerance levels) involved in different steps of a method through simulations. We compare our method with two representative methods of \cite{MadanYor} and \cite{BallottaKyriakou} (hereinafter abbreviated as MY and BK, respectively) which entail various approximations. We abbreviate the version of our time change decomposition method incorporating the double CFTP scheme (see Section \ref{subsec:perfectsimGGC}) as TCD and the version incorporating the approximation scheme (see Section \ref{subsubsec:APPscheme}) as TCD-app. Because approximations are inevitable in the infinite-variation case while exact simulation methods are available in the finite-variation case, in the following, we shall only consider the infinite-variation CGMY model.

\subsection{Prerequisites}
To better understand the following simulation results, one needs know more details about the involved specific errors in different steps of the two existing methods MY and BK. First, recall that the simulation method MY of \cite{MadanYor} relies on constructing the CGMY time change by shaving (using the rejection method of \cite{Rosinski}) the approximate $Y/2$-stable process, which is, moreover, built on truncating jumps with sizes below certain threshold (i.e., the MY's $\varepsilon$, adopting the same notation of \cite{MadanYor}) of a $Y/2$-stable process. Some comments related to the errors involved in this method are listed as follows.
\begin{itemize}
\item[(\textbf{MY.i}).] The MY's $\varepsilon$ (an \emph{error parameter}) is determined by controlling a Berry-Esseen-type upper bound estimate (see Theorem 3.1 of \cite{AsmussenRosinski}) for the distance between the target and approximate $Y/2$-stable distribution functions to be less than a pre-specified tolerance level, say, 1\%.
\item[(\textbf{MY.ii}).] The shaving (or the rejection sampling) step of MY method relies on the evaluation of a truncation function (see equation (18) of \cite{MadanYor}).
\item[(\textbf{MY.iii}).] However, both the above Berry-Esseen-type upper bound and truncation function do not admit closed-form expressions as functions of the model and error parameters. Hence, solving for the optimal (largest possible) MY's $\varepsilon$ and evaluating the truncation function must rely on numerical procedures, whose computing costs can be substantial (see the discussion in the last paragraph on page 40 of \cite{MadanYor}). For simulation scenarios with \emph{a fixed set of model parameters}, pre-computation and -tabulation are possible to save computing time, but for simulation-based model calibrations, this method of saving computing costs does not apply.
\item[(\textbf{MY.iv}).] Most importantly, we do not know how the above errors translate explicitly into simulation biases measured by, e.g., the $L^1$ distance between the target and approximate CGMY increment. That is, there is no closed-form expression which relates the bound of this distance explicitly to the error parameter, i.e., the MY's $\varepsilon$. See also the discussions in the Introduction of \cite{BallottaKyriakou}. Therefore, given a pre-specified tolerance level on simulation biases, we do not know the optimal choice of the MY's $\varepsilon$.
\end{itemize}
Second and similarly, some concerns about the specific errors involved in the method BK of \cite{BallottaKyriakou} are listed below.
\begin{itemize}
\item[(\textbf{BK.i}).] The method BK involves regularization error, truncation error and discretization error with error parameters $D$ (determines the truncation of the domain of the distribution function), $L$ (determines the truncation of the domain of the Fourier transform) and $N$ (determines the discretization spacing for the discrete Fourier transform), respectively, adopting the notation of \cite{BallottaKyriakou}.
\item[(\textbf{BK.ii}).] The bounds of the above three errors generally do not admit closed-from expressions as functions of the model and error parameters. Hence, searching the optimal choices of error parameters given a pre-specified error tolerance level relies on numerical procedures. For simulation scenarios with a fixed set of model parameters, pre-computation and -tabulation are possible and help reduce computing burden (see Table II of \cite{BallottaKyriakou}). However, computational costs induced by these numerical procedures in simulation-based model calibrations can be immense.
\item[(\textbf{BK.iii}).] Most importantly, we do not have a closed-form expression about the relation between the bound of simulation bias and that of the above specific errors. Therefore, we have no guide that helps determine the optimal choices of the BK's $D$, $L$ and $N$ given a pre-specified tolerance level on simulation biases.
\end{itemize}

The aim of this study is to demonstrate the relevance of having an explicit guide on the optimal choices of error parameters given a pre-specified tolerance level on a simulation bias measure such as the $L^1$ distance between the target and approximate CGMY increment. The above points (MY.iv) and (BK.iii) clearly show that methods MY and BK lack such explicit guides while our method does not as can be seen from the discussions in Section \ref{subsec:TransInterpError}.

Because we are not pursuing  optimal encoding either of our method or  of the other methods (in fact, we simply adopt the C++ codes for the MY method available on Peter Tankov's personal website\footnotemark\footnotetext{The URL for Peter Tankov's website is: http://www.proba.jussieu.fr/pageperso/tankov/}  and translate the algorithm of the BK method straightforwardly into C++) and more importantly, the encodings of these methods do not take the numerical procedures described in points (MY.iii) (where one simply sets an ad hoc value for the MY's $\varepsilon$) and (BK.ii) into account, the comparison of computing speeds among different methods is somewhat inappropriate and should be interpreted carefully, although these codes have been implemented in the same computing environment. The simulation experiments are performed on a desktop PC with an Intel$^\circledR$  Core\texttrademark \ i5-8400T CPU @ 1.70\,GHz 1.70\,GHz and 8.00\,GB RAM. All programs are coded in the C++ programming language and compiled by Microsoft Visual Studio 2010.

\subsection{Simulation Results}\label{subsec:simresults}
We are now ready to present the details of our simulation study. The set of model parameters used is as follows: $C=0.42$, $G=4.37$, $M=191.2$ and $Y=1.0102$, which are chosen by taking the estimation results from calibrating the model to the option price data, with IBM being the underlying asset in Table 3 on page 327 of \cite{CGMY2002} as reference. Without loss of generality and for ease of exposition, we compare the performances of four methods (i.e., MY, BK, TCD and TCD-app) in simulation-based estimation of the mean of $X(t)$, where $t=1/52$ year (or a week). In this case, the true mean is easily obtained as
$$
{\rm E}(X(t))=-tC\Gamma(1-Y)\left(G^{Y-1}-M^{Y-1}\right)=-0.0317757,
$$
which facilitates the evaluation of different simulation methods. Suppose we generate $B$ i.i.d. samples $(\widetilde{X}(t)_i)_{1\leq i\leq B}$ using one of the four simulation methods, then an estimator $\widehat{{\rm E}}(X(t))$ of ${\rm E}(X(t))$ is given by the following sample mean
\begin{align}\label{eq:EstMeanCGMY}
\widehat{{\rm E}}(X(t)):=\frac{1}{B}\sum_{i=1}^B\widetilde{X}(t)_i,
\end{align}
whose estimation error consists of both the sampling error $({\rm Var}(\widetilde{X}(t))/B)^{1/2}$ and the bias induced by various approximations involved in the simulation method.\footnotemark\footnotetext{Similar to the mean of $X(t)$, its variance also admits a closed-form formula as ${\rm Var}(X(t))=tC\Gamma(2-Y)(G^{Y-2}+M^{Y-2})$. For our method, since ${\rm Var}(\widetilde{X}(t))<{\rm Var}(X(t))$, the sampling error is bounded by $({\rm Var}(X(t))/B)^{1/2}\approx0.0004$ under the setting of model parameters in this paper and when $B=10,000$.} We use $\widetilde{X}(t)$ as a generic notation for the variate generated by one of the four simulation methods. $\widetilde{X}(t)$ is close in distribution to $X(t)$. The sampling error can be made arbitrarily small by increasing the number of Monte Carlo trials $B$ and estimated by $(\widehat{{\rm Var}}(\widetilde{X}(t))/B)^{1/2}$, where
\begin{align}\label{eq:EstSampError}
\widehat{{\rm Var}}(\widetilde{X}(t)):=\frac{1}{B}\sum_{i=1}^B\left(\widetilde{X}(t)_i-\widehat{{\rm E}}(X(t))\right)^2.
\end{align}
However, increasing the number of Monte Carlo trials $B$ does not help reduce the bias. We set $B=10,000.$

For the method MY, we report the simulation-based estimation results across different choices of the MY's $\varepsilon$ (i.e., the jump truncation threshold). For the method BK, mean estimates are produced across different choices for the tolerance level (i.e., the BK's $\varepsilon$) of the sum of three errors, i.e., regularization error, truncation error and discretization error. Notice that the optimal choices of the BK's $D$, $L$ and $N$ given different choices of the BK's $\varepsilon$ for two particular sets of model parameters are pre-computed and -tabulated in Table II of \cite{BallottaKyriakou}. However, the two methods do not provide explicit guides on the choices of right (optimal) error parameters (i.e., the MY's $\varepsilon$, the BK's $\varepsilon$ or the BK's $D$, $L$ and $N$) given a tolerance level on the aforementioned simulation bias.

As to our methods, from Section \ref{subsec:TransInterpError} (see (\ref{eq:L2boundCGMYAppError}) in particular) and the discussion immediately following (\ref{eq:DirichletAppErrBound}), one can easily see that our tolerance levels on specific errors, i.e., $\varepsilon$ and $\tilde{\varepsilon}$, exactly control the magnitude of the above-mentioned simulation bias. Hence, given a pre-specified tolerance level on this bias, we know the optimal (largest possible) choices of $\varepsilon$ and $\tilde{\varepsilon}$. We only report the estimation results of our methods for $\varepsilon=\tau\tilde{\varepsilon}=10^{-5}$, which gives biases that have an order of magnitude about $10^{-3}$ (see the following Remark \ref{rem:CommOnMeanEstErr}), our target level.

\begin{remark}\label{rem:CommOnMeanEstErr}
In simulation-based estimation of ${\rm E}(X(t))$ using our method, a more detailed analysis on the estimation bias than that given in (\ref{eq:L2boundCGMYAppError}) and the discussion immediately following (\ref{eq:DirichletAppErrBound}) can be done. Take the approximating error $X_2(t)=\theta\epsilon_L(t)+(\epsilon_L(t))^{1/2}W_2(1)$ in Section \ref{subsec:TransInterpError} for example, bias is only due to $\theta\epsilon_L(t)$ since $(\epsilon_L(t))^{1/2}W_2(1)$ has mean zero. Bias due to $\theta\epsilon_L(t)$ is bounded by $|\theta|\varepsilon\approx0.0009$ (see (\ref{eq:L2boundCGMYAppError})) under the setting of this simulation study. In the same simulation setting, $(\epsilon_L(t))^{1/2}W_2(1)$ yields a sampling error bounded by $(\varepsilon/B)^{1/2}\approx$3.162e-05 which is negligible compared with the bias. However, in general (e.g., in estimating ${\rm Var}(X(t))$), $(\epsilon_L(t))^{1/2}W_2(1)$ may lead to bias.
\end{remark}

The simulation results are summarized in Table 1, on the basis of which we make the following comments:
\begin{itemize}
\item Most importantly, from the aforementioned discussion, we know \emph{in advance} that an optimal choice of $\varepsilon=\tau\tilde{\varepsilon}=10^{-5}$ leads to biases with a target order of magnitude about $10^{-3}$. That is, given a pre-specified error tolerance level, we can set values for $\varepsilon$ and $\tilde{\varepsilon}$ on purpose rather than at random, avoiding either large simulation biases or extra computing costs. The orders of magnitudes of biases and sampling errors given in Remark \ref{rem:CommOnMeanEstErr} (see also footnote 4) are consistent with the sampling errors and estimation errors reported in Panel B of Table 1.
\item By contrast, the two methods under comparison do not provide explicit guides on the optimal choices of MY's $\varepsilon$ and BK's $\varepsilon$ given a pre-specified tolerance level on simulation biases. One may either choose the MY's/BK's $\varepsilon$ at random or perform pre-computation and -tabulation, which is time-consuming, as in Table 1. From Panel A of Table 1, on the one hand, for a wide range of choices of the MY's/BK's $\varepsilon$ (from $10^{-3}$ to as small as $10^{-12}$), our methods outperform the MY and BK methods in terms of estimation error under the studied model. On the other hand, as the MY's/BK's $\varepsilon$ decreases, computing times of the MY and BK methods increase. Therefore, choosing the MY's/BK's $\varepsilon$ at random runs the risk of leading to either large biases or extra computing costs.
\item From the computing times reported in Panel B of Table 1, one can see that the approximation scheme TCD-app (with computing time 2.878 seconds) substantially reduces the computational burden without virtual loss of estimation accuracy (in terms of sampling error and estimation error) compared with the TCD method (with computing time 133.999 seconds). For the MY and BK methods to achieve the same level of estimation accuracy as our methods, smaller MY's/BK's $\varepsilon$ than that in Table 1 should be used, but this would lead to larger computing costs. Notice that the computing time of the MY method for $\varepsilon=10^{-12}$ is already 141.541 seconds which is even larger than that of the TCD method. The BK method with $\varepsilon=10^{-14}$ (unreported in Table 1) can achieve roughly the same estimation accuracy as our methods, having sampling error 0.00041484 and estimation error -0.0005505, but it then consumes longer computing time (21.744 seconds) than our TCD-app method.
\end{itemize}

\begin{center}
+++ Insert Table 1 about here +++
\end{center}

\begin{remark}
Although we believe that the differences among the computing complexities of  the different methods will eventually become insignificant as a result of  advances in information technology, a more efficiently designed algorithm for the proposed method is still of practical importance at present. Because an exact path simulation method is not available for CGMY processes of infinite variation, a method that has both transparently interpretable approximation error(s) and an efficiently designed algorithm is desirable. However, as can be seen from Sections \ref{subsec:perfectsimGGC}--\ref{subsubsec:APPscheme}, the construction of a simulation algorithm for our method with uniformly bounded complexity over the whole parameter space is not a straightforward task, and we leave this as a topic for future research.
\end{remark}

\section{CONCLUDING REMARKS}\label{sec:conclusion}
We have found a new and easy-to-implement path simulation method for CGMY processes   with either finite or  infinite variation. Our method is based on a time change representation of the CGMY process and a decomposition of its time change into a \emph{finite} \emph{generalized gamma convolution} subordinator and an independent error term. In the infinite-variation case, in contrast to  the existing path simulation methods of \cite{MadanYor} and \cite{BallottaKyriakou}, which entail various nontrivial specific approximation errors that are difficult to quantify in, e.g., derivatives pricing applications, our proposed method is more appealing in that its approximation errors have a more transparent interpretation, i.e., the upper bound of the $L^1$ distance between the approximate variable and the target CGMY increment admits closed-form expression as a function of the pre-specified tolerance levels ($\varepsilon$ and $\tilde{\varepsilon}$) on specific errors, see Section \ref{subsec:TransInterpError} and the discussion immediately following (\ref{eq:DirichletAppErrBound}). This facilitates the choice of the right (optimal) error tolerance levels, avoiding either large simulation biases or extra computing costs. Simulation results support the above findings showing that our method is advantageous over the methods of \cite{MadanYor} and \cite{BallottaKyriakou} under the studied model.

\section{ACKNOWLEDGMENTS}
We are very grateful to the Editor-in-Chief Professor Ming Hu, an Associate Editor and two anonymous referees for their valuable comments and constructive suggestions that lead to improvements of the paper. The idea behind this work originates from a conversation between Professor Lancelot F. James and the second author. Zhiyuan Zhang's research is supported by the National Nature Science Foundation of China (71301097 and 91546202).

\appendix{}
\section{Algorithms}
\subsection{The Double Rejection Sampler when $0<Y<1$}\label{subsec:appdoublerejection}

Before we introduce the double rejection method, we need the following notation.

Recall that $\lambda:=tC\Gamma(1-Y)/Y$ as in Section \ref{subsec:exactmethodYzeroone}. Define $\gamma:=M^Y\lambda Y(1-Y)$, $\xi:=\pi^{-1}[(2+(\pi/2)^{1/2})(2\gamma)^{1/2}+1]$, $\psi:=\pi^{-1}\exp(-\gamma\pi^2/8)(2+(\pi/2)^{1/2})(\gamma\pi)^{1/2}$, $w_1:=\xi(\pi/(2\gamma))^{1/2}$, $w_2:=2\psi\pi^{1/2}$, $w_3:=\xi\pi$, and $b:=(1-Y)/Y$. The Zolotarev function $A(u)$ is defined as
$$
A(u):=\left[\frac{(\sin(Yu))^Y(\sin((1-Y)u))^{1-Y}}{\sin(u)}\right]^{\frac{1}{1-Y}},~~~0\leq u\leq\pi.
$$
Moreover, define $B(x):=A(x)^{-(1-Y)}$, $0\leq x\leq\pi$, and $B(0):=\lim_{x\downarrow0}B(x)=Y^{-Y}(1-Y)^{-(1-Y)}$. The algorithm for generating a random number from the distribution of $X^+(t)$ is  as follows:

\begin{center}
\begin{small}
\begin{itemize}
\item[] \texttt{repeat} \texttt{repeat} \texttt{generate $V$ and $W'$ uniformly on $[0,1]$}\\
            \texttt{~~~~~~~~~~~~~if $\gamma\geq1$ then if $V<\frac{w_1}{w_1+w_2}$ then $U\leftarrow|N|/\gamma^{1/2}$ where $N\sim$ Normal(0,1)\\
            ~~~~~~~~~~~~~~~~~~~~~~~~~~~else $U\leftarrow\pi(1-W'^2)$\\
            ~~~~~~~~~~~~~else if $V<\frac{w_3}{w_2+w_3}$ then $U\leftarrow\pi W'$\\
            ~~~~~~~~~~~~~~~~~~else $U\leftarrow\pi(1-W'^2)$\\
            ~~~~~~~~~~~~~generate $\widetilde{W}$ uniformly on $[0,1]$\\
            ~~~~~~~~~~~~~let $\zeta=(B(U)/B(0))^{1/2}$, $\phi=(\gamma^{1/2}+Y\zeta)^{1/Y}$, $z=\phi/\left(\phi-\gamma^{1/(2Y)}\right)$,\\
            ~~~~~~~~~~~~~and $\rho=\frac{\pi e^{-M^Y\lambda\left(1-\zeta^{-2}\right)}\left(\xi e^{-\frac{\gamma U^2}{2}}I\{U\geq0,\gamma\geq1\}+\frac{\psi}{(\pi-U)^{1/2}}I\{0<U<\pi\}+\xi I\{0\leq U\leq\pi,\gamma<1\}\right)}{(1+(\pi/2)^{1/2})\gamma^{1/2}/\zeta+z}$}
    \item[] \texttt{~~~~~~until $U<\pi$ and $Z:=\widetilde{W}\rho\leq1$}
    \item[] \texttt{~~~~~~let $a=A(U)$, $m=(bM\lambda^{1/Y}/a)^Y$, $\delta=(mY/a)^{1/2}$, $a_1=\delta(\pi/2)^{1/2}$, $a_2=\delta$\\
    ~~~~~~$a_3=z/a,$ $s=a_1+a_2+a_3$}\\
    \texttt{~~~~~~generate $V'$ uniformly on $[0,1]$\\
    ~~~~~~if $V'<a_1/s$ then generate $N'\sim$ Normal(0,1) and let $X'\leftarrow m-\delta|N'|$\\
    ~~~~~~else if $V'<(a_1+a_2)/s$ then generate $X'$ uniformly on $[m,m+\delta]$\\
    ~~~~~~~~~~~else generate $E'\sim$ Exponential$(1)$ and let $X'\leftarrow m+\delta+E'a_3$\\
    ~~~~~~let $E=-\log(Z)$}
\item[] \texttt{until $X'\geq0$ and $a(X'-m)+M\lambda^{1/Y}(X'^{-b}-m^{-b})-\frac{N'^2}{2}I\{X'<m\}-E'I\{X'>m+\delta\}\leq E$}
\item[] \texttt{return $\lambda^{1/Y}/X'^b$}
\end{itemize}
\end{small}
\end{center}

\subsection{The Double CFTP Sampler}\label{subsec:doublecftp}
We present the double CFTP algorithm (cf. \cite{JZ:2011}) for generating random numbers from the distribution of $\mathcal{D}$ defined through (\ref{eq:srochasticequationgeneral}) in Section \ref{subsec:perfectsimGGC}. Recall that the density function $h(\cdot)$ of $B$ is bounded from below on $[0,1]$ by a constant $c_h>0$ and $0<Q\leq c_Q<\infty.$ Let $(U_i)_{i\geq1}$ be ${\rm Uniform}[0,1]$ random variables and $Q$ and $Q'$ have the same distribution. The algorithm consists of the following steps (a)--(d):

\begin{center}
\begin{small}
\begin{itemize}
\item[(a)] \texttt{For $i=-1,-2,\ldots$: }\\
\texttt{keep generating $(U_i,Q_i,Q'_i)$ and
store $(Q_i,Q'_i)$ }\\
\item[] \texttt{until $U_\mathbb{T}\leq
|Q_\mathbb{T}-Q'_\mathbb{T}|c_h/(2c_Q)$ ;}
\item[(b)] \texttt{Keep $\mathbb{T}$ and set $\mathcal{D}=Q_\mathbb{T}\wedge
Q'_\mathbb{T}+2c_QU_\mathbb{T}/c_h$;}
\item[(c)] \texttt{For $i=\mathbb{T}+1,\mathbb{T}+2,\ldots,-1$, iterate the following}:\\
\texttt{repeat generate $U'\sim$ Uniform$[0,1]$, $\xi_{1/2}\sim$ Bernoulli$(1/2)$ and $B$, and set $X'=(1-B)\mathcal{D}+BQ_i\xi_{1/2}+BQ'_i(1-\xi_{1/2})$}
\item[] \texttt{until: $$U'\left[h\left(\frac{X'-\mathcal{D}}{Q_i-\mathcal{D}}\right)\frac{1}{|Q_i-\mathcal{D}|}
    +h\left(\frac{X'-\mathcal{D}}{Q'_i-\mathcal{D}}\right)\frac{1}{|Q'_i-\mathcal{D}|}\right]> c_h/c_Q$$ or $X'< Q_i\wedge Q'_i$ or $X'> Q_i\vee Q'_i$,}\\
    \texttt{then update $\mathcal{D}=X'$;}
\item[(d)] \texttt{Return $\mathcal{D}$.}
\end{itemize}
\end{small}
\end{center}

\begin{landscape}
\begin{table}[H]\renewcommand{\arraystretch}{1.2}\footnotesize
    \vspace{-0.4in}
    \caption{{ Simulation results for estimating the mean of the CGMY increment $X(t)$ using (23) based on different simulation methods. Notice that given a pre-specified tolerance level on the estimation error (or bias), we know \emph{in advance} the optimal choices for the error parameters (TCD's $\varepsilon$ and $\widetilde{\varepsilon}$) of our method while this is not the case for the methods MY and BK, where no closed-form expression like (14) in Section 4.4 is available for the bound of the estimation error (or bias) as a function of the model and error parameters (MY's/BK's $\varepsilon$). Our TCD-app method outperforms methods MY and BK in terms of both estimation accuracy and computing speed, see the text for more discussions regarding the results in this table.}}
    \vspace{-0.08in}
    \begin{center}

\begin{tabular}{c@{\hspace{0.2cm}}c@{\hspace{0.2cm}}c@{\hspace{0.2cm}}c@{\hspace{0.2cm}}c@{\hspace{0.2cm}}c@{\hspace{0.2cm}}c@{\hspace{0.2cm}}c@{\hspace{0.2cm}}c@{\hspace{0.2cm}}c}
\hline\hline
 \textbf{Panel A} &   &\multicolumn{4}{c}{MY}& \multicolumn{4}{c}{BK} \\
\cmidrule(lr){3-6}
\cmidrule(lr){7-10}
   \vspace{-0.08in}
MY's/BK's  & True  & Estimated   &   & Estimation  & Comp.  & Estimated   &   & Estimation  &Comp.     \\
$\varepsilon$ & mean & mean  & s.e. & error & time   & mean  & s.e. & error &time      \\
 \hline
 $10^{-3}$& -0.0317757 &  -0.1225280 & 0.00766599  & 0.0907521 & 0.032  & -0.0264359  & 0.001286430  & -0.0053398 & 0.005    \\ \hline
 $10^{-4}$& -0.0317757 &  -0.0955166 & 0.00819276  & 0.0637409 & 0.038  & -0.0295841  & 0.000818592 & -0.0021916 & 0.008    \\ \hline
 $10^{-5}$& -0.0317757 &  -0.0766466 & 0.00521004  & 0.0448709 & 0.081  & -0.0343991  & 0.000541411  & 0.0026234 &  0.011   \\ \hline
 $10^{-6}$& -0.0317757 &  -0.0848639 & 0.00759333  & 0.0530882 & 0.216  & -0.0333892  & 0.000495514  & -0.0016135 & 0.013    \\ \hline
 $10^{-7}$& -0.0317757 &  -0.0656537 & 0.00170570  & 0.0338780 & 0.616  & -0.0283489  & 0.000387015  & -0.0034268 & 0.018    \\ \hline
 $10^{-8}$& -0.0317757 &  -0.0679307 & 0.00177919  & 0.0361550 & 1.875  & -0.0339155  & 0.000465107  & 0.0021398 &  0.032   \\ \hline
 $10^{-9}$& -0.0317757 &  -0.0631211 & 0.00164765  & 0.0313454 & 5.808  & -0.0332689  & 0.000450621  & -0.0014932 & 0.056    \\ \hline
 $10^{-12}$&-0.0317757  & -0.0220309 & 0.00031954  & -0.0097448 & 141.541  & -0.0328860  &  0.000465153 & 0.0011103 & 2.184    \\ \hline
 \textbf{Panel B}  & &\multicolumn{4}{c}{TCD}&\multicolumn{4}{c}{TCD-app ($\tau\widetilde{\varepsilon}=\varepsilon$)}\\
\cmidrule(lr){3-6}
\cmidrule(lr){7-10}
   \vspace{-0.08in}
TCD's  & True  & Estimated   &   & Estimation  & Comp.  & Estimated   &   & Estimation  &Comp.    \\
$\varepsilon$ & mean & mean  & s.e. & error & time   & mean  & s.e. & error &time     \\
 \hline
   $10^{-5}$&-0.0317757  &-0.0313361   &0.00045346   & \textbf{-0.0004396} & \textbf{133.999}  & -0.0314646  & 0.00045613  & \textbf{-0.0003111} &  \textbf{2.878}  \\ \hline
\end{tabular}
\end{center}
\vspace{-0.1in}
{\scriptsize Note. Model parameters are set as $C=0.42$, $G=4.37$, $M=191.2$, $Y=1.0102$ and $t=1/52$. The number of Monte Carlo trials is set as $B=10^4$ such that the order of sampling error ($10^{-4}$, see either footnote 4 or the values reported in the s.e. columns of Panel B) is smaller than that of the bias ($10^{-3}$, the target level, see Remark 2) for our method with $\varepsilon=\tau\widetilde{\varepsilon}=10^{-5}$. The ``true mean'' of $X(t)$ is computed as in the beginning of Section 6.2. The ``estimated mean'' is defined by (23). The ``s.e.'' column reports the estimated sampling errors defined based on (24). The ``estimation error'' is given by the difference between the ``true mean'' and ``estimated mean'', consisting of both sampling error and bias. Computing times (``Comp. time'' column) are measured in seconds. MY and BK refer to the methods of $[$23$]$ and $[$3$]$, respectively. TCD and TCD-app refer to our methods with the exact double CFTP and approximation schemes, respectively.}
\label{table:DesignIEuropeanCall}
\end{table}
\end{landscape}

\end{document}